\documentclass[aps,prd,preprint,superscriptaddress,showpacs,floatfix,nobibnotes]{revtex4-1}

\usepackage{latexsym}
\usepackage{amsmath}
\usepackage{amssymb}
\usepackage{graphicx}

\usepackage{bm}
\usepackage{epsfig}
\usepackage{subfigure}
\newcommand{\bea}{\begin{eqnarray}}
\newcommand{\eea}{\end{eqnarray}}

\begin{document}

\title{The 2PI effective action at four loop order in $\varphi^4$ theory}

\author{M.E. Carrington}
\email[]{carrington@brandonu.ca} \affiliation{Department of Physics, Brandon University, Brandon, Manitoba, R7A 6A9 Canada}\affiliation{Winnipeg Institute for Theoretical Physics, Winnipeg, Manitoba}

\author{B.A. Meggison}
\email[]{brett.meggison@gmail.com} \affiliation{Department of Physics, Brandon University, Brandon, Manitoba, R7A 6A9 Canada}\affiliation{Winnipeg Institute for Theoretical Physics, Winnipeg, Manitoba}

\author{D. Pickering}
\email[]{pickering@brandonu.ca} \affiliation{Department of Mathematics, Brandon University, Brandon, Manitoba, R7A 6A9 Canada}

\date{\today}

\begin{abstract}

It is well known that perturbative pressure calculations show poor convergence.
Calculations using a two particle irreducible (2PI) effective action show improved convergence at the 3 loop level, but no calculations have been done at 4 loops. 
We consider the 2PI effective theory for a symmetric scalar theory with quartic coupling in 4-dimensions.
We calculate the pressure and two different non-perturbative vertices as functions of coupling and temperature. 
Our results show that the 4 loop contribution can become larger than the 3 loop term when the coupling is large. 
This indicates a breakdown of the 2PI approach, and the need for higher order $n$PI approximations.
In addition, our results demonstrate the renormalizability of 2PI calculations at the 4 loop level. 
This is interesting because the counterterm structure of the 2PI theory at 4 loops is different from the structure at $n\le 3$ loops. 
Two vertex counterterms are required at the 4 loop level, but not at lower loop order. This unique feature of the 2PI theory has not previously been verified numerically.

\end{abstract}


\normalsize
\maketitle

\normalsize

\section{Introduction}
\label{introduction-section}

There are many interesting systems which involve non-perturbative physics.
Problems of this kind cannot be solved by expanding in some small parameter. 
One possible technique is the use of $n$-particle irreducible ($n$PI) effective theories \cite{Jackiw1974,Norton1975,berges-hierarchy}.
The basic motivation is the hope that they can
be applied to nonabelian gauge theories, but there has been little progress to date in this direction. 
Calculations are complicated by issues with gauge fixing \cite{Smit2003,Zaraket2004} and renormalizability \cite{vanHees2002,Blaizot2003,Serreau2005,Serreau2010}. 

In this paper we work with the simplest $n$PI theory, which is the 2PI version (also known as the $\Phi$-derivable approximation). 
One of the first successful uses of the 2PI theory was a calculation of entropy in QCD \cite{Blaizot1999}.
It can also be used to study transport coefficients in scalar theories \cite{Aarts2004} and QED \cite{Carrington2006}, and the approach to equilibrium in far from equilibrium systems \cite{berges-cox,berges-aarts,berges-baier,tranberg-smit,tranberg-aarts,tranberg-laurie}. 
The gauge dependence of the QED pressure at 2 loops was studied in \cite{Borsanyi-2pi}.
Phase transitions in the $SU(N)$ Higgs theory were studied at the 3 loop level in \cite{guy-2014}.
We also note that other methods exist for the calculation of purely thermodynamic quantities in non-perturbative systems. 
One of the most successful is screened perturbation theory, which has been applied to scalar theories 
\cite{andersen-scalar1,andersen-scalar2,Braaten2001,andersen-scalar3}, QED \cite{Strickland-2pi} and QCD \cite{strickland-qcd1,strickland-qcd2}.

In this paper we study an equilibrium symmetric $\varphi^4$ theory, and work at 4 loop order in the 2PI theory. 
The primary goal of this work is to study the convergence of the skeleton expansion. 
Calculations were done at the 3 loop level in \cite{Berges2005a}, and improved convergence properties were found, relative to perturbative calculations. 
We find that the 4 loop approximation agrees well with the 3 loop one when the coupling constant is not too large, but as the coupling grows 4 loop contributions become important. This indicates that in a situation where non-perturbative physics is important, higher order $n$PI approximations may be needed.

In addition, our calculation is interesting because it provides numerical verification of the renormalizability of the 2PI theory at the 4 loop level. 
The renormalization of the symmetric 2PI theory requires, in general, two different coupling constant counterterms which must be determined from two renormalization conditions that are imposed on different 4-point functions \cite{Serreau2005,Serreau2010}.
However, at the 2 loop and 3 loop levels the structure is much less complicated - only one counterterm is required. 
Our calculation thus provides a non-trivial check of the renormalizability of the 2PI effective theory. 

This paper is organized as follows. 
In section \ref{2PI-section} we review the 2PI formalism.
We describe the numerical method in section \ref{numerical-section} (more details can be found in Refs. \cite{Berges2005a,Fu2013,Fu2014,Fu2015}).
In section \ref{results-section} we present our results, and we conclude in section \ref{conclusions-section}.



\section{The 2PI Effective Theory}
\label{2PI-section} \vspace{5pt}

In this section we review some definitions and techniques used in 2PI calculations. 
In most equations in this paper we suppress integrals and the arguments that denote the space-time dependence of functions. As an example of this notation, the quadratic term in the action is written:
\bea
\frac{i}{2}\int d^4 x\,d^4 y\,\varphi(x)G_{\rm no\cdot int}^{-1}(x-y)\varphi(y) ~~\longrightarrow~~\frac{i}{2}\varphi\, G_{\rm no\cdot int}^{-1}\varphi\,.
\eea

\subsection{Action}

The classical action is 
\begin{eqnarray}
\label{action}
&& S[\varphi] =\frac{i}{2}\varphi \,G_{\rm no\cdot int}^{-1}\varphi -\frac{i}{4!}\lambda_b\varphi^{4}\,,~~
iG_{\rm no\cdot int}^{-1} = -(\Box + m_b^2)\,.
\eea
For notational convenience we use a scaled version of the physical coupling constant.
The extra factor of $i$ will be removed when rotating to Euclidean space to do numerical calculations. 
The effective action is obtained in the standard way.
We use a BPHZ renormalization procedure and write all expressions in terms of renormalized quantities. 
The effective action can be written generically as 
\bea
\label{gamma-division}
&& \Gamma[\phi,G] =\Gamma_{\rm no\cdot int}[\phi,G] +\Gamma_{\mathrm{int}}[\phi,G]\,.
\eea
We define $i\Gamma[G] = \Phi[G]$, $i\Gamma_{\rm no\cdot int}[G] = \Phi_{\rm no\cdot int}[G]$ and $i\Gamma_{\rm int}[G] = \Phi_{\rm int}[G]$.
We work to order $\lambda^3$ in the skeleton expansion. 
The non-interacting part of equation (\ref{gamma-division}) is
\bea
\label{gamma-no-int}
\Gamma_{\rm no\cdot int}[G] = \frac{i}{2}\phi G^{-1}_{\rm no\cdot int}\phi  + \frac{i}{2}\mathrm{Tr}\ln G^{-1}
+\frac{i}{2}\mathrm{Tr}G_{\rm no\cdot int}^{-1}G\,.
\eea

The interacting piece can be divided into terms that do and do not contain counterterms. 
The counterterm contributions are (see Fig. \ref{ct-fig})
\bea
\label{L-ct}
\Phi_{\rm int\cdot ct}&& = -\frac{i}{2}(\delta Z_2\Box+\delta m_2^2)\phi^2
-\frac{i}{2}(\delta Z_0\Box+\delta m_0^2) {\rm Tr}G
+ \frac{1}{4!} \delta\lambda_4 \phi^{4}
+ \frac{1}{4}\delta\lambda_{\rm tp} \phi^{2}G \nonumber\\
&& +\frac{1}{3}\lambda\,\delta\lambda_{\rm egg}\phi^2 G^3
 +\frac{1}{8} \delta\lambda_{\rm et} G^2  
+\frac{1}{24} \delta\lambda_{\rm bb} \lambda G^3 +{\cal O}(\lambda^4)\,.
\eea
In the exact theory, all counterterms of the same type are equal (for example, all vertex counterterms are equal: $\delta\lambda_4$ = $\delta\lambda_{\rm tp}$ = $\delta\lambda_{\rm egg}$ = $\delta\lambda_{\rm et}$ = $\delta\lambda_{\rm bb}$). At a finite order of truncation, the different counterterms in equation (\ref{L-ct}) could in principle be defined differently.
\begin{figure}
\begin{center}
\includegraphics[width=14cm]{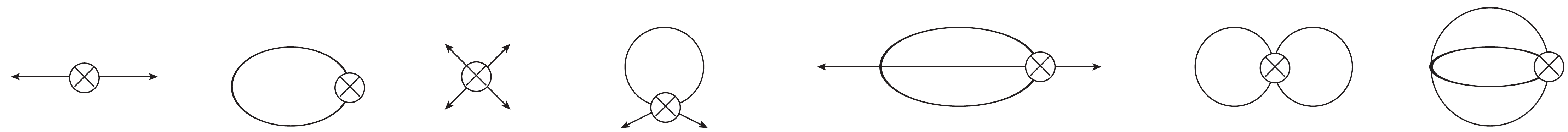}
\end{center}
\caption{Contributions to $\Phi_{\rm ct}$ to order $\lambda^3$. The diagrams represent the terms in Eq. (\ref{L-ct}) in the order they appear in the equation.  \label{ct-fig}}
\end{figure}
The non-counterterm contributions to $\Phi_{\rm int}$ are represented as 
\bea
\label{phi-int-no-ct}
\Phi_{\rm int\cdot no\cdot ct} = \frac{1}{4!}\lambda \phi^4+ \frac{1}{8}\lambda G^2+\frac{1}{6}\lambda^2 G^3 +\frac{1}{48}\lambda^2 G^3+\frac{1}{8}\lambda^3 G^5+\frac{1}{48}\lambda^3 G^6\,.
\eea 
In the symmetric theory the only loop diagrams that contribute are the second, fourth and sixth terms in (\ref{phi-int-no-ct}), which are shown in Fig. \ref{phi-fig}.
\begin{figure}[htb]
\begin{center}
\includegraphics[width=10cm]{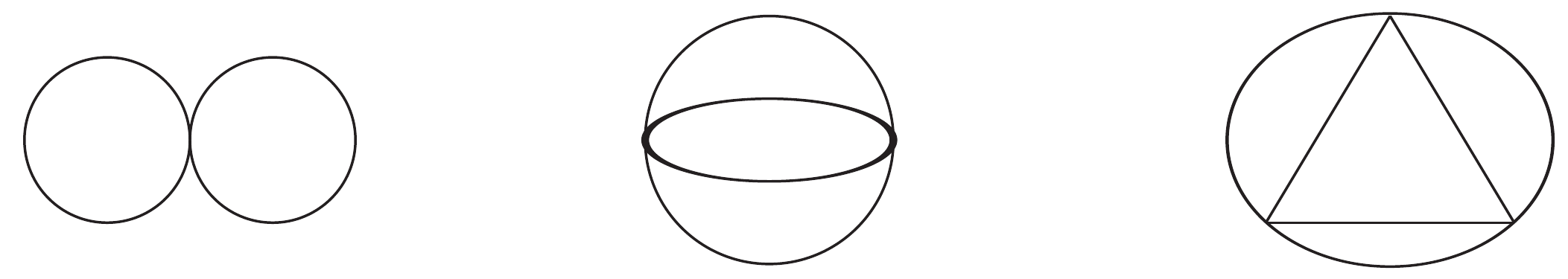}
\end{center}
\caption{The second, fourth and sixth terms in (\ref{phi-int-no-ct}).\label{phi-fig}}
\end{figure}

We define the kernels:
\bea
\label{kernels}
\Phi^{(n\,m)}[\tilde G] = 2^m\frac{\delta^{n+m}}{\delta\phi^n\delta G^m}\Phi_{\rm int}[\phi,G]\bigg|_{\substack{\phi=0\\G=\tilde G}}\,.
\eea
These kernels appear in the self-consistent integral equations that generate the non-perturbative $n$ point functions of the theory. 

\subsection{Integral equations}

The stationary condition is 
\bea
\label{stat-cond}
\frac{\delta \Phi[\phi,G]}{\delta G}\bigg|_{\substack{\phi=0\\G=\tilde G}}=0\,.\\
\eea
This equation takes the form
\begin{subequations}
\label{all-2}
\begin{equation}
\label{dyson}
\tilde G^{-1} = G_{\rm no\cdot int}^{-1}-\Sigma[\tilde G]\,,\\
\end{equation}
where $\Sigma$ is the kernel $\Phi^{(0\,1)}$ defined in (\ref{kernels}):
\begin{equation}
\label{kernel2}
\Sigma[\tilde G] = 2\frac{\delta \Phi_{\rm int}[\phi,G]}{\delta G}\bigg|_{\substack{\phi=0\\G=\tilde G}}\,.
\end{equation}
\end{subequations}
Equation (\ref{all-2}) is a self-consistent equation for the propagator $\tilde G$ which has the structure of a Dyson equation.

For any $n$PI theory one can also derive integral equations which produce non-perturbative vertices that resum diagrams in specific channels \cite{Carrington-BS-2013,Russell2013}. In our calculation there are two such vertices, which we call $M$ and $V$. These vertices are obtained from the self-consistent equations
\begin{subequations}
\label{all-4}
\bea
\label{M-defn}
&& M[\tilde G] = \Lambda[\tilde G] + \frac{1}{2}  \Lambda[\tilde G]\; \tilde G^2 \, M[\tilde G] \,,\\
\label{V-defn}
&& V[\tilde G] = \lambda+\delta\lambda_4 + 3\big(M[\tilde G] - \Lambda[\tilde G]) \,,
\eea
where $\Lambda$ is the 4-kernel $\Phi^{(0\,2)}$ obtained from (\ref{kernels})
\begin{equation}
\label{kernel4}
\Lambda[\tilde G] = 4\frac{\delta^{2} \Phi_{\rm int}[\phi,G]}{\delta G^2}\bigg|_{\substack{\phi=0\\G=\tilde G}}\,.
\end{equation}
\end{subequations}
We comment on the physical content of equation (\ref{all-4}), which is somewhat obscured by the notation we are using. The vertex $M$, which is usually called the Bethe-Salpeter vertex, resumms the kernel $\Lambda$ in the $s$ channel. The vertex $V$ involves a resummation in all three ($s$, $t$ and $u$) channels. Using our shorthand notation which suppresses indices, the three channels are not shown separately, but combine to produce the factor (3) in equation (\ref{V-defn}).

The goal is to solve the self-consistent integral equations (\ref{all-2}) and (\ref{all-4}). 
It appears that these equations are not coupled, and that we could first solve (\ref{all-2}) for the propagator $\tilde G$, and then use the result and solve (\ref{all-4}) for the vertices $M$ and $V$. In fact, we will see below that the two integral equations are coupled, because of the counterterm structure. Once the counterterms have been determined, 
the two equations decouple, and finite temperature calculations are therefore easier. \\



From this point on we suppress the tilde on the self-consistent propagator and write simply $G$. 
We work in Euclidean space. 
We use an obvious shorthand notation in which functional dependence on four independent momentum components is represented as a single capital letter. When the four momentum is zero, we again use only one argument. For example, $\Lambda(p_4,p_1,p_2,p_3,k_4,k_1,k_2,k_3) \to \Lambda(P,K)$, $G(0,0,0,0) \to G(0)$, etc.
The equation for the 4-kernel (from (\ref{kernel4})) is
\bea
\label{Lam-euc}
&& \Lambda(P,K) \approx -\lambda -\delta\tilde\lambda + \lambda(\lambda+2\delta\tilde\lambda)\int dL G(L)G(L+P+K) \\
&& - \lambda^3\int dL \int dS\,G(S)G(L)\big[G_{S+L+P}G_{L+P-K}+G_{S+L-P}G_{L-P-K} + \frac{1}{2} G_{L+P+K}G_{S+P+K}\big]\,,\nonumber
\eea
where we have used a shorthand notation for the propagators that depend on three momenta to save space (for example $G(S+P+K) = G_{S+P+K}$).
The kernel $\Lambda$ contains counterterms from the sixth and seventh diagrams in Fig. \ref{ct-fig} (the reason they are denoted with tildes as $\delta\tilde\lambda$  will be explained below). Note that the expression for $\Lambda$ in (\ref{Lam-euc}) does not come directly from (\ref{kernel4}). The full $\Lambda$ contains contributions from $t$ and $u$ channels which can be written as 2 times the $t$ channel piece when the kernel is embedded in the BS equation (\ref{M-euc}), by shifting dummy variables. This symmetrization has already been done in (\ref{Lam-euc}), and this is indicated by the wiggly equal sign. 
The counterterm $\delta\tilde\lambda$ will be determined from the renormalization condition 
\bea
\label{Lambda-rc}
\Lambda(0,0)=-\lambda\,.
\eea
We rewrite this renormalization condition as follows
\begin{subequations}
\label{deltatildelambda}
\begin{equation}
\label{deltatildelambda-1}
\Lambda(P,K) = -\delta\tilde\lambda + \Lambda_d(P,K)\,,
\end{equation}
\begin{equation}
\label{deltatildelambda-2}
\delta\tilde\lambda = \lambda+\Lambda_d(0,0) \,,
\end{equation}
\begin{equation}
\label{deltatildelambda-3}
\Lambda(P,K) = -\lambda + \big[\Lambda_d(P,K)-\Lambda_d(0,0)\big]\,.
\end{equation}
\end{subequations}
Equation (\ref{deltatildelambda}) is a self-consistent equation for $\delta\tilde\lambda$, since $\Lambda_d(P,K)$ is a function of $\delta\tilde\lambda$ (see equation (\ref{Lam-euc})). 
It is straightforward to show that the quantity in square brackets in equation (\ref{deltatildelambda-3}) is finite.

The BS equation (\ref{M-defn}) in momentum space is  
\bea
\label{M-euc}
&& M(P,0) = -\Delta\lambda+\Lambda(P,0)+\frac{1}{2}\int dQ \,\big[-\Delta\lambda+\Lambda(P,Q)\big]\,G^2(Q)\,M(Q,0)\,.
\eea
We note that since this equation resums only the $s$ channel, one can fix the momentum on one side of the vertex $M$. 
The new counterterm $\Delta\lambda$ is an addition contribution to the kernel of the BS equation from the sixth diagram in Fig. \ref{ct-fig}.
It is determined from the renormalization condition 
\bea
\label{M-rc}
M(0,0)=-\lambda
\eea 
which, together with (\ref{deltatildelambda}), gives a self consistent equation for $\Delta\lambda$ of the form
\bea
\label{Deltalambda}
&& \Delta\lambda = \frac{1}{2}\int dQ \,\big[-\Delta\lambda+\Lambda(0,Q)\big]\,G^2(Q)\,M(Q,0)\,.
\eea
Notice that (\ref{deltatildelambda}) and (\ref{Deltalambda}) are coupled, since $\Lambda(0,Q)$ depends on $\delta\tilde\lambda$. 
To understand the role of the counterterm $\Delta\lambda$, we imagine expanding the BS equation instead of solving it self-consistently. The kernels $\Lambda$, which were made finite with the counterterm $\delta\tilde\lambda$, are chained together in the $s$ channel. 
The 2PI nature of the kernels guarantees that no new divergences are generated, except in the loops that join the kernels together. These divergences are cancelled by the counterterm $\Delta\lambda$. 
The vertex $V$ in equation (\ref{V-defn}) is finite when $M$ and $\Lambda$ are, and therefore we can set $\delta\lambda_4$ to zero.

Next we consider the 2-point function which is obtained from 
equation (\ref{all-2}), including counterterm diagrams of the form shown in the second, sixth and seventh parts of Fig \ref{ct-fig}.
The vertex counterterm is the sum of the two counterterms calculated above:
\bea
\label{deltalambda}
\delta\lambda = \delta\tilde\lambda+\Delta\lambda\,.
\eea
The resulting equations are
\bea
\label{sigma-euc}
\Sigma(P) &&=  \delta m^2+\delta Z P^2 +\frac{(\lambda+\delta\lambda)}{2}\int dQ G(Q)\nonumber\\
&&  - \frac{1}{6}\lambda(\lambda+2\delta\lambda)\int dQ\int dL G(L)G(L+Q)G(P+Q)\nonumber\\
&& +\frac{\lambda^3}{4}\int dS\int dL \int dM\,G(S)G(L)G(S+M)G(L+M)G(P-M)\,,\\[4mm]
\label{G-euc}
G(P)&& = \big(P^2 + m^2 +\Sigma(P)\big)^{-1}\,.
\eea
The counterterms $\delta Z$ and $\delta m^2$ are obtained from the usual renormalization conditions
\bea
\label{renormconds-euc}
&& G^{-1}(0) =m^2\,,~~ \frac{d}{dP^2}G^{-1}\bigg|_{P=0} =1\,.
\eea
For later use we define the quantity $\Sigma_d(P)$
\bea
\Sigma(P) = \delta m^2+\delta Z P^2+\Sigma_d(P)\,.
\eea

\subsection{Comparision with 3 loop 2PI theory}

At this point it is easy to see that calculations in the 2PI theory are considerably simpler when the effective action is truncated at the 3 loop level. The reason is that  
the 4-point kernel has only a global divergence at this order. 
Equations (\ref{Lam-euc}) and (\ref{deltatildelambda-1}) give
\bea
&& \Lambda_d(P,K)\big|_{\rm 3~loops} =  -\lambda  + \lambda^2\int dL G(L)G(L+P+K) \,,
\eea
and from (\ref{deltatildelambda-2}) we see that the equation that determines $\delta\tilde\lambda$ is not a self-consistent equation at 3 loop order. The result is that the two counterterms $\Delta\lambda$ and $\delta\tilde\lambda$ can be immediately combined as in (\ref{deltalambda}), and the BS equation can be written so that it depends on only one coupling constant counterterm, which can be determined from (\ref{M-rc}). Schematically we have
\bea
&& {\rm kernel}_{\rm 4\,loops} = -\Delta\lambda-\delta\tilde\lambda + \Lambda_d^{(4)}[\tilde\delta\lambda] \,,\\
&& {\rm kernel}_{\rm 3\,loops} = -\Delta\lambda-\delta\tilde\lambda + \Lambda_d^{(3)} = -\delta\lambda + \Lambda^{(3)}_d \,.
\eea

\subsection{Pressure}

The pressure can be obtained from the effective action using 
\bea
P=\frac{T}{V}\Phi
\eea
where $V$ is the 3-volume. We include all contributions to $\Phi$ from equations (\ref{gamma-no-int}), (\ref{L-ct}) and (\ref{phi-int-no-ct}).
\bea
\label{P0-def}
&& P_0 = - \frac{1}{2}\int dQ \ln G^{-1}_{\rm no\cdot int}(Q) \to \frac{\pi^2 T^4}{90} \\
&& P_1= - \frac{1}{2}\int dQ \ln \big[G^{-1}(Q)\,G_{\rm no\cdot int}(Q)\big] -\frac{1}{2}\int dQ \,\big[G^{-1}_{\rm no\cdot int}(Q) G(Q)-1\big]\\
&& P_2 =  -\frac{1}{2} \int dQ\,(Q^2 \delta Z+\delta m^2)\, G(Q) \\
&& P_3 = -\frac{1}{8}(\lambda+\delta\lambda)\int dQ\,G(Q)\int dL\, G(L) \\
&& P_4  = \frac{1}{48}\lambda(\lambda+2\delta\lambda) \int dS\int dL\int dQ G(S)G(L)G(Q)G(S+L+Q)\\
&& P_5  = -\frac{1}{48} \lambda^3 \int dQ\big[\int dS\,G(S)G(S+Q)\, \int dL\,G(L)G(L+Q)\, \int dM\,G(M)G(M+Q) \big]\nonumber\\ \\
&& P_{\rm sum}=P_0+P_1+P_2+P_3+P_4+P_5\,.
\eea
There is an overall temperature independent divergence that can be removed by a `cosmological constant' renormalization, which means requiring that the vacuum pressure be zero:
\bea
\label{P-final}
\Delta P = P_{\rm sum} - P_{\rm sum}(T=0)\,.
\eea
The arrow on the right side of (\ref{P0-def}) indicates that a temperature independent constant has been dropped. This constant would be removed by the shift in (\ref{P-final}) anyway.
The term $P_0$ is the non-interacting ($\lambda=0$) pressure. 
We want to compare $\Delta P$ to the non-interacting expression, so we define
\bea
P = \frac{\Delta P}{P_0}\,.
\eea

\section{Numerical method}
\label{numerical-section}

We want to solve the integral equations (\ref{Lam-euc}), (\ref{M-euc}), (\ref{sigma-euc}) and (\ref{G-euc}). The counterterms are determined from (\ref{Lambda-rc}), (\ref{M-rc}), (\ref{deltalambda}) and (\ref{renormconds-euc}).
We use always $m=1$, which means we give all dimensionful quantities in mass units. 
In order to do the numerical calculation, we restrict to a box in co-ordinate space of finite volume $L^3 \beta$. Fourier transforming to momentum space one obtains discrete frequencies and momenta. This can be written
\bea
&& \int \frac{dp_4}{2\pi} \prod_{i=1}^3 \int^\infty_{-\infty}\frac{dp_i}{2\pi} f(p_4,p_i)
\to \frac{m_t m_s^3}{(2\pi)^4} \sum_{n_{4}=-\frac{N_t}{2}+1}^{\frac{N_t}{2}} \prod_{i=1}^3 \sum_{n_i=-\frac{N_s}{2}+1}^{\frac{N_s}{2}} f(m_t n_4,m_s n_i) \,,\\[4mm]
\label{mhat-defn}
&& m_t = 2\pi T = \frac{2\pi}{N_t a_t}\,,~~m_s= 2\pi L^{-1} = \frac{2\pi}{N_s a_s}\,,~~
L=a_s N_s\,,~~T=\frac{1}{a_t N_t}\,.
\eea
The parameters $a_t$ and $a_s$ are the lattice spacing in the temporal and spatial directions. Indices which fall outside of the range $\{-N/2+1,N/2\}$ are wrapped inside using periodic boundary conditions. 

It is well known that the scalar $\phi^4$ theory in 4-dimensions is non-interacting if it is considered
as a fundamental theory valid for arbitrarily high momentum scales (quantum triviality), but the renormalized coupling is non-zero if the theory has an ultraviolet cutoff and an infrared regulator. In our calculation the mass $m$ regulates the infrared and the lattice spacing parameter provides an ultraviolet cutoff. 

There are certain restrictions on the values that can be chosen for the paramters $a_t$, $a_s$, $N_t$ and $N_s$, which are discussed below. 
We have checked that results are independent of the choices of these parameters, within these restrictions. 
We use lattice spacing $a_t=a_s=1/12$ and in the spatial direction we use $N_s=32$.
The renormalization is done with $N_t=128$.
We have verified numerically that the corresponding temperature gives the zero temperature limit, and we refer to it from here on as zero temperature. 
Finite temperature calculations are obtained from $126\ge N_t\ge 6$. 

The numerical method replaces a continuous integration variable with infinite limits by a discrete sum over a finite number of terms. 
For numerical accuracy, we need that the upper limit of the sum is big and the step size is small. This means we require
$P_{\rm max}\sim \frac{1}{a_s} \gg 1$ and $\Delta P \sim \frac{1}{L}=\frac{1}{N a_s} \ll 1$.
The number of lattice points $N$ is limited by memory and computation time, and therefore there is a limit on how small $a_s$ can be taken while maintaining $N a_s$ big. 
However, there is another more subtle issue that limits how small we can choose $a_s$.
The theory has a Landau pole at a scale that decreases when $\lambda$ increases. 
When $\lambda$ becomes large, $a_s$ must increase ($P_{\rm max}$ must decrease) so that the integrals are cut off in the ultraviolet at a scale below the Landau scale. 
However, decreasing the ultraviolet cutoff $P_{\rm max}$ will eventually cause important contributions from the momentum phase space to be missed. 
When $\lambda$ has increased to the point that the Landau scale has moved down and dipped into the momentum regime over which the integrand is large, physically meaningful results cannot be obtained. 
In our calculation we have determined that the maximum coupling we can calculate is $\lambda\approx 8$ (using $a_s=1/12$ and $N_s=32$). 

We use an iterative relaxation method to solve the self-consistent equations. 
In the equations below, an index in round brackets indicates the iteration number of a given quantity. We start with the bare propagator and the BS vertex obtained from the renormalized 4-kernel:
\bea
&& G^{(0)}(P)  = G_{{\rm no}\cdot{\rm int}}(P) = \big[P^2+m^2\big]^{-1}\\[4mm]
&& \Lambda^{(0)}_d(P,Q) = \Lambda[G^{(0)},\delta\tilde\lambda=0]\\
&& \delta\tilde\lambda^{(0)} = \lambda+\Lambda_d^{(0)}(0, 0)\\
&& \Lambda^{(0)}(P,Q) = -\delta\tilde\lambda^{(0)} + \Lambda_d^{(0)}(P,Q)\\[4mm]
&& M^{(0)}(Q,0) = \Lambda^{(0)}(Q,0)\,.
\eea
At the first iteration we update the propagator using
\bea
&& \Sigma_d^{(1)}(P) = \Sigma_d[G^{(0)},\delta\lambda^{(0)}] \\[2mm]
&& \delta m^{2(1)} = -\Sigma_d^{(1)}(0) \label{cond-delm}\\[2mm]
&& \delta Z^{(1)} = -\frac{1}{m_s^2}\big(\Sigma_d^{(1)}(0,0,0,1)-\Sigma_d^{(1)}(0,0,0,0)\big)\label{cond-delZ}\\[2mm]
&& \Sigma^{(1)}(P) = \delta m^{2(1)} + \delta Z^{(1)}P^2 +\Sigma_d^{(1)}(P) \\[2mm]
&& G^{(1)}(P) = \big[\big(G^{(0)}(P)\big)^{-1}+\Sigma^{(1)}(P)\big]^{-1}\,.
\eea
Using this updated propagator we calculate the updated 4-kernel and BS vertex:
\bea
&& \Lambda_d^{(1)}(P,Q) = \Lambda_d[G^{(1)},\delta\tilde\lambda^{(0)}]\\[2mm]
&& \delta\tilde\lambda^{(1)} = \lambda+\Lambda_d^{(1)}(0,0)\label{cond-dellamtilde}\\[2mm]
&& \Lambda^{(1)}(P,Q) = -\delta\tilde\lambda^{(1)} + \Lambda_d^{(1)}(P,Q)\\[2mm]
&& M^{(1)}(P,0) = \big(-\Delta\lambda^{(0)} + \Lambda^{(1)}(P,0)\big) \nonumber\\
&&~~~+\frac{1}{2}\int dQ \big(-\Delta\lambda^{(0)}+\Lambda^{(1)}(P,Q)\big)\big(G^{(1)}(Q)\big)^2M^{(0)}(Q,0)\\[2mm]
&& \Delta\lambda^{(1)} = \Delta\lambda^{(0)} + \lambda + M^{(1)}(0,0)\\[2mm]
&& \delta\lambda^{(1)} = \delta\tilde\lambda^{(1)}+\Delta\lambda^{(1)}\,.
\eea
Continuing in the same fashion, the quantities obtained from the first iteration are used to obtain the second iteration results. 
Iterations are terminated when the relative maximum difference between the $(i+1)$th iteration and the $i$th, for any quantity, at any point in momentum space, is less than $10^{-4}$. 


\section{results}
\label{results-section}

We compare results from a truncation in the skeleton expansion at 2, 3 and 4 loops.
We will use circles (blue), diamonds (green) and boxes (red) as markers to represent truncation at 2, 3 and 4 loops.
On graphs that show both of the 4-vertices $M$ and $V$, we use open symbols for the vertex $M$ and solid symbols for $V$. 

In Fig. \ref{MVversusg} we show the zero momentum BS vertex $M(0,0,0,0)$ and symmetric vertex $V(0,0,0,0)$ at fixed temperature as a function of $g=\sqrt{\lambda/24}$ (which would correspond to an interaction term in the Lagrangian of the form $\frac{\lambda}{24}\varphi^4 = g^2\varphi^4$). 
Agreement is good between all levels of truncation when $g$ is small, as expected. The 4 loop contributions become large as $g$ increases.

In Fig. \ref{MVversusT} we show the two vertices at fixed $\lambda$ as functions of the temperature. At zero temperature they are renormalized to the chosen value of the coupling. Deviations between different orders in the approximation are evident as the temperature increases.

In Fig. \ref{PversusL} we show the pressure as a function of $g=\sqrt{\lambda/24}$ at $T=2$. The well known oscillations that appear in the perturbative calculation are not present. However, as the coupling grows the 4 loop result deviates increasingly from the 3 loop one. 

Fig \ref{Mlogas} demonstrates that the renormalization is done correctly. We reduce the lattice spacing in the spatial direction ($a_s$) while holding the length of the box ($L=a_s N$) fixed. The graph shows $-V(0,0,0,0)$ versus $\log(1/a_s)$ for $\lambda=2$ and $T=1$. For comparision we show the curve that results when the renormalization is done incorrectly, using the 3 loop approximation but including an additional vertex counterterm on the 3 loop basketball diagram. When the renormalization is done correctly, the curve is almost completely flat. 

\begin{figure}[htb]
\begin{minipage}{8cm}
\center
\includegraphics[width=1.0\textwidth]{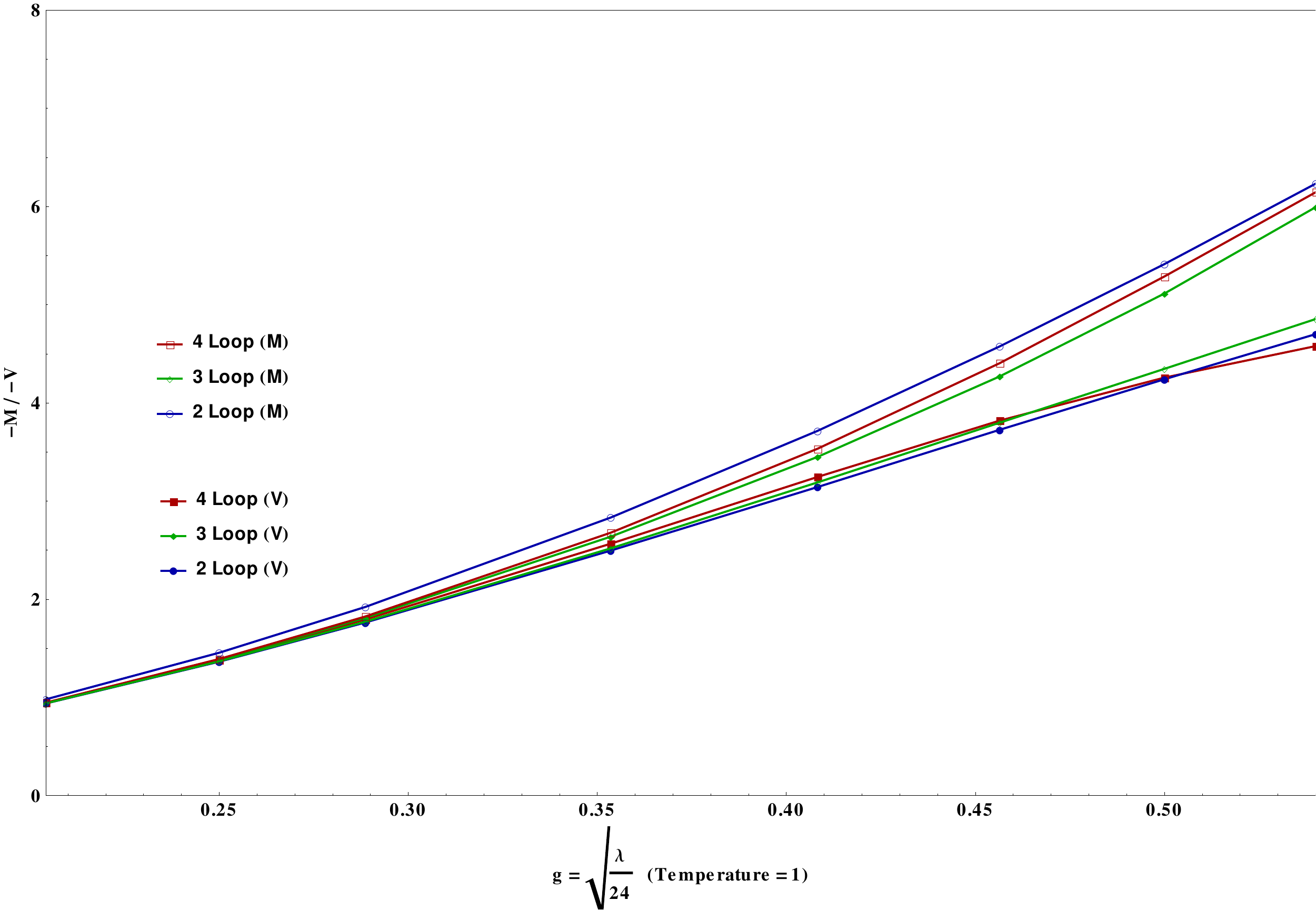}
\end{minipage}
\begin{minipage}{8cm}
\center
\includegraphics[width=1.0\textwidth]{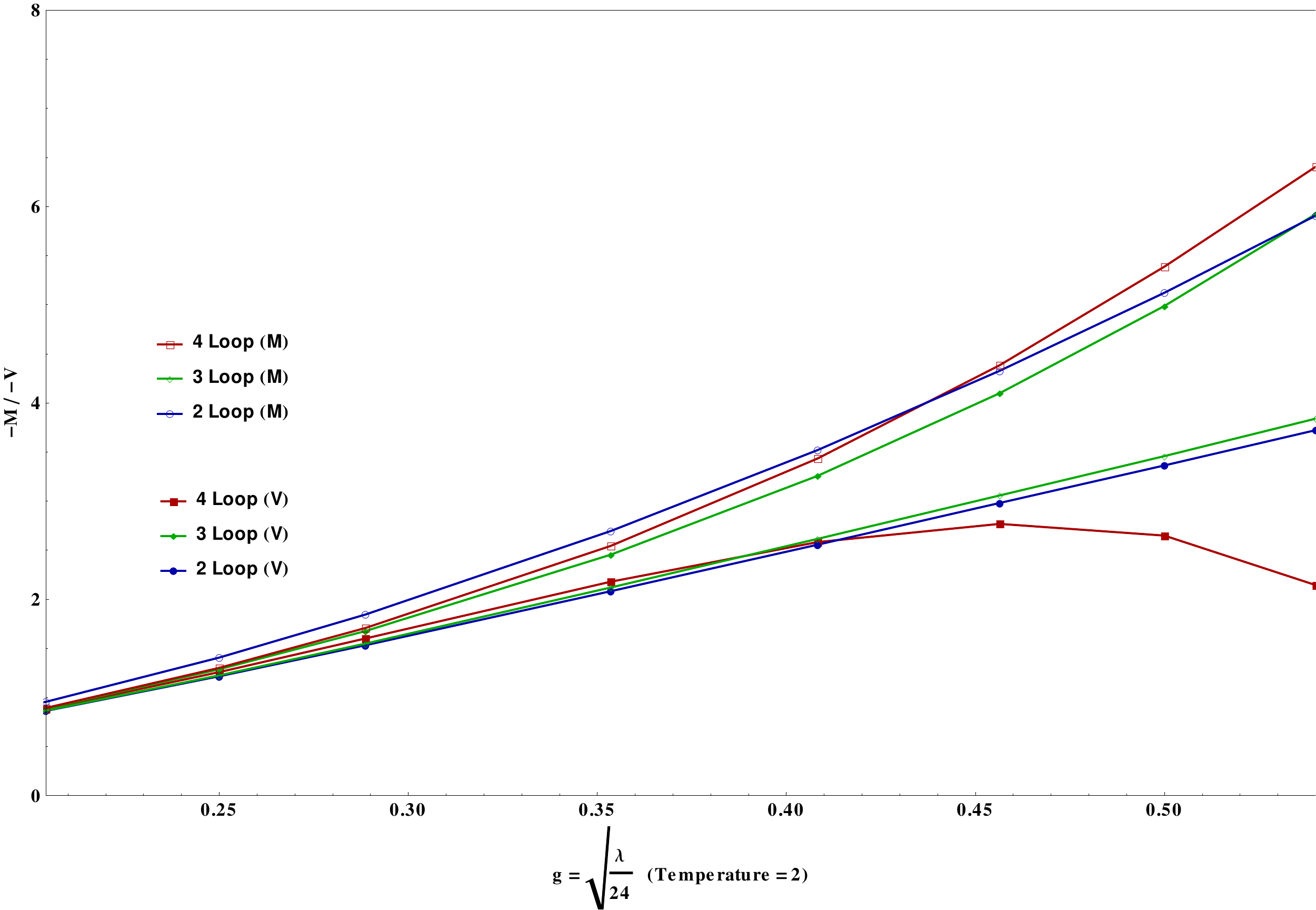}
\end{minipage}
\caption{The vertices $M$ and $V$ versus $g=\sqrt{\lambda/24}$ for two different values of temperature.}
\label{MVversusg}
\end{figure}

\begin{figure}[htb]
\begin{minipage}{8cm}
\center
\includegraphics[width=1.0\textwidth]{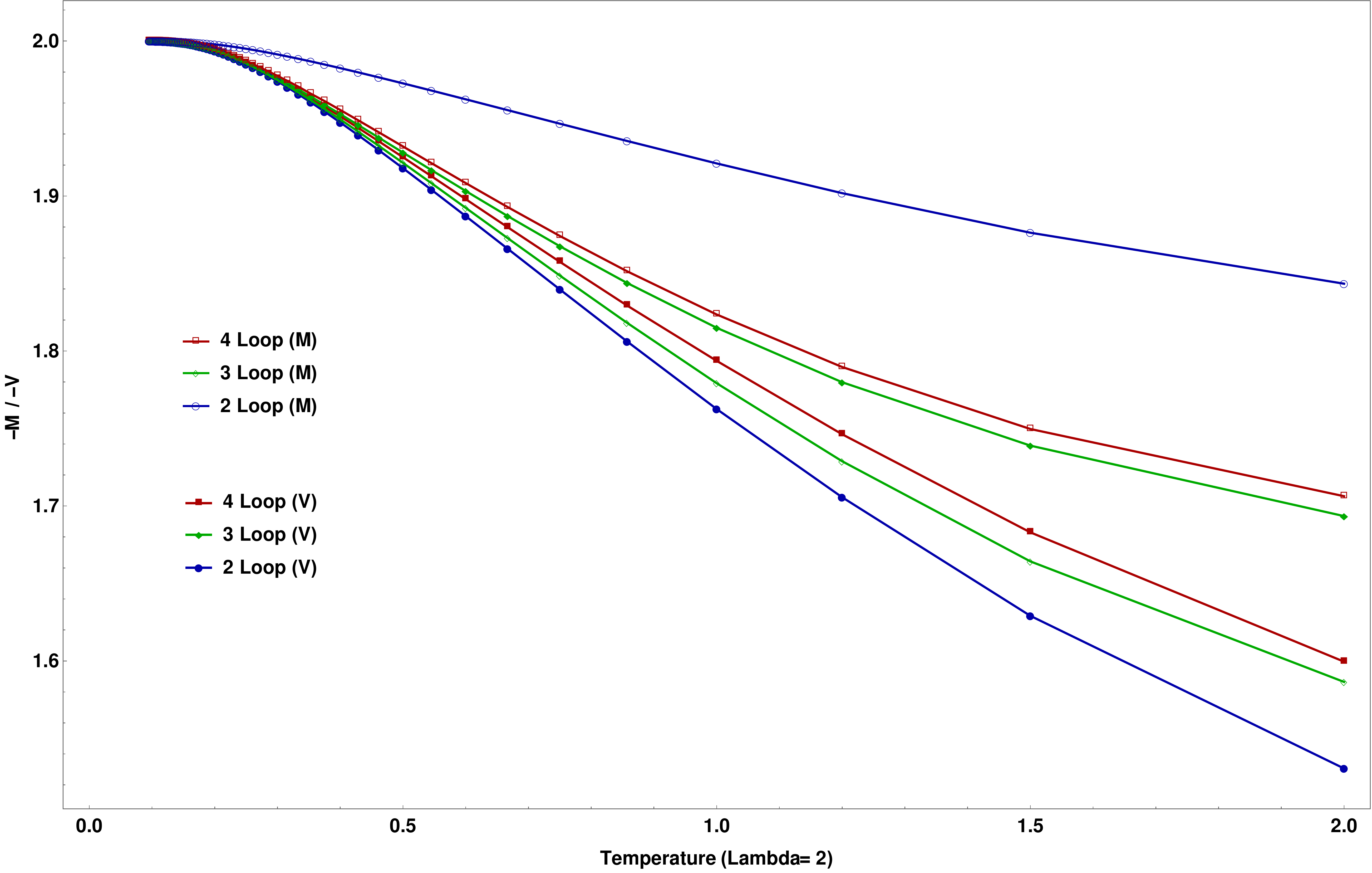}
\end{minipage}
\begin{minipage}{8cm}
\center
\includegraphics[width=1.0\textwidth]{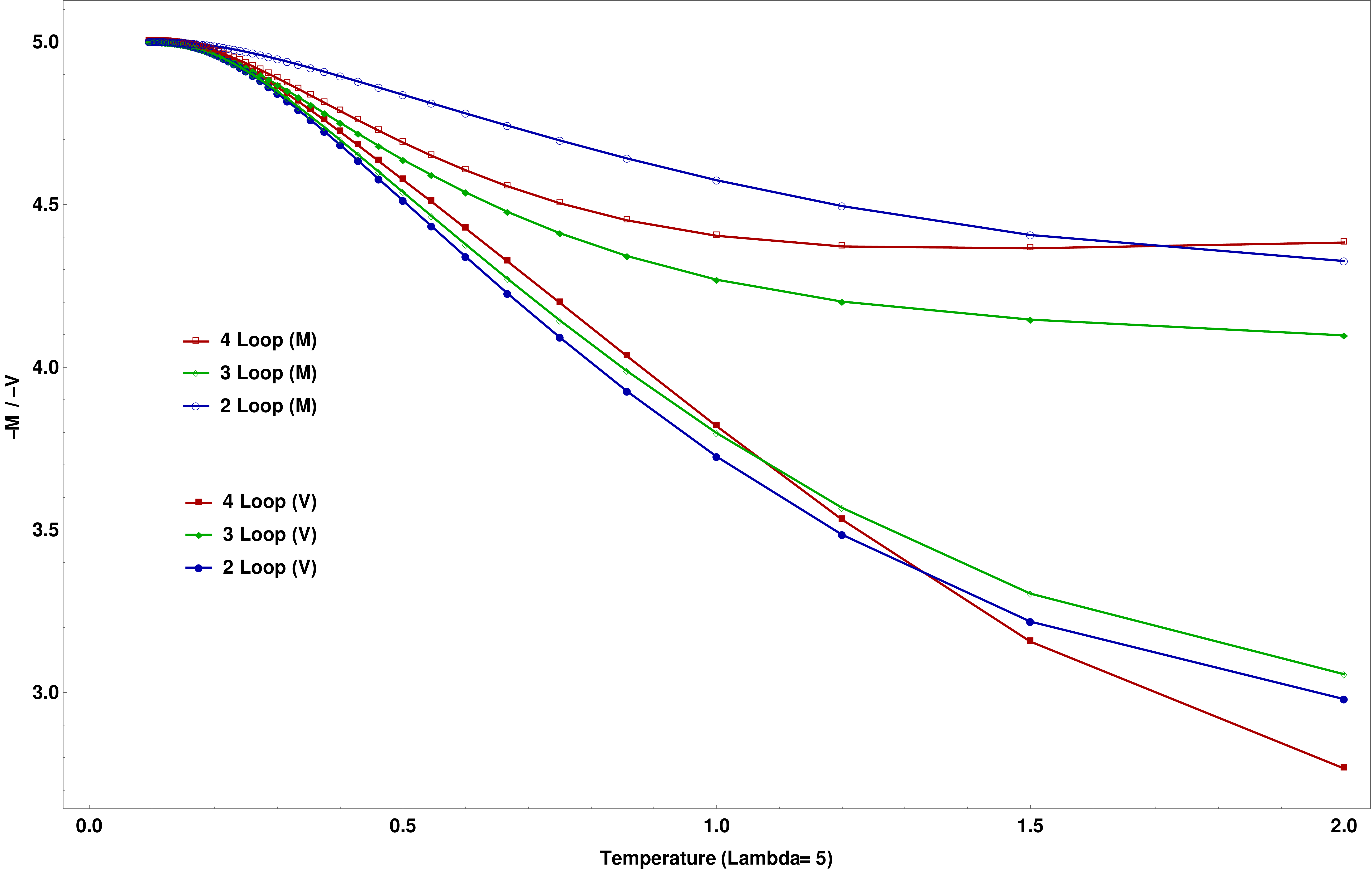}
\end{minipage}
\caption{The vertices $M$ and $V$ versus temperature for two different values of $\lambda$.}
\label{MVversusT}
\end{figure}

\begin{figure}[htb]
\begin{center}
\includegraphics[width=11cm]{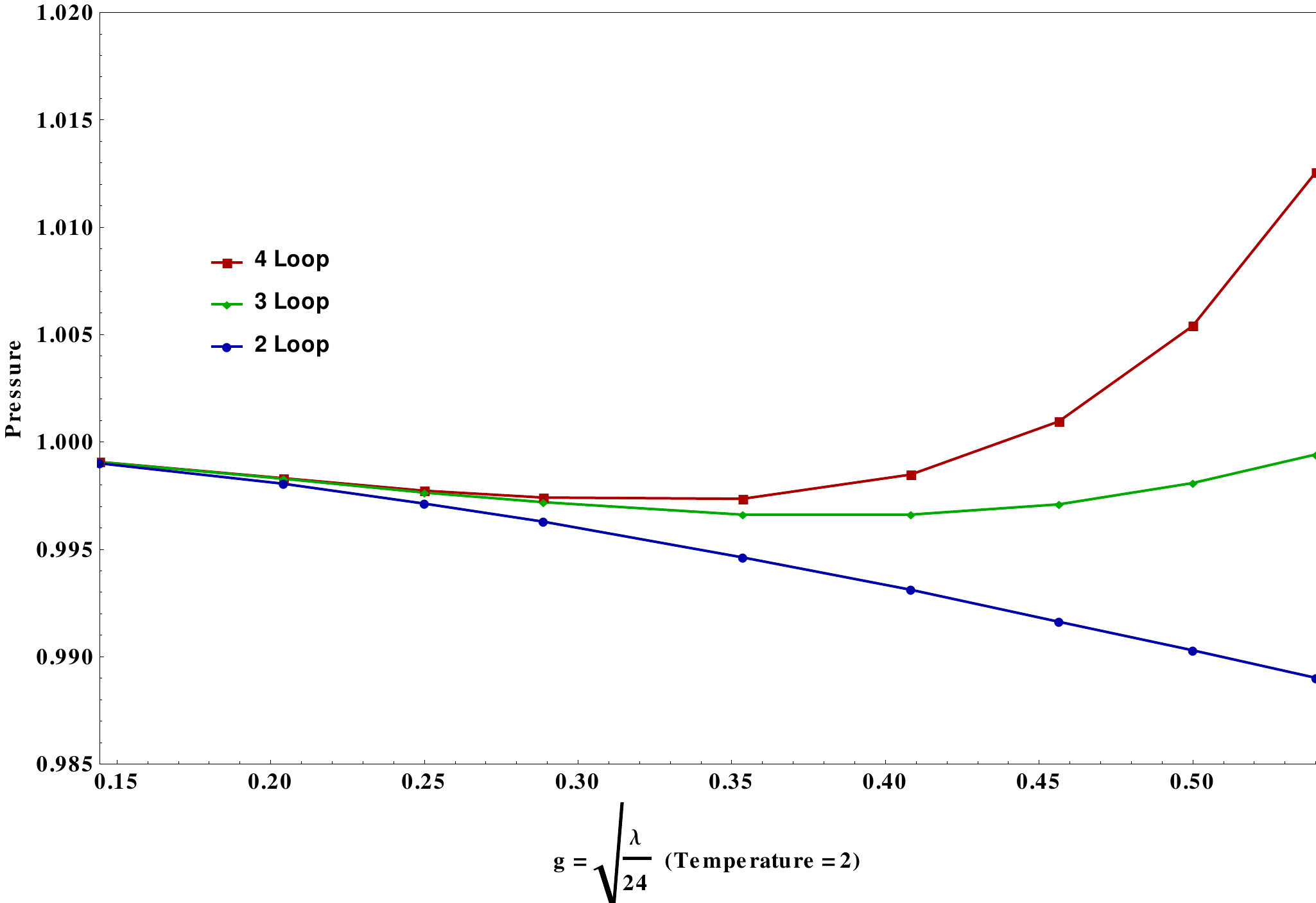}
\end{center}
\caption{Pressure versus $g=\sqrt{\lambda/24}$ at $T=2$. }
\label{PversusL}
\end{figure}
\begin{figure}[h]
\begin{center}
\includegraphics[width=11cm]{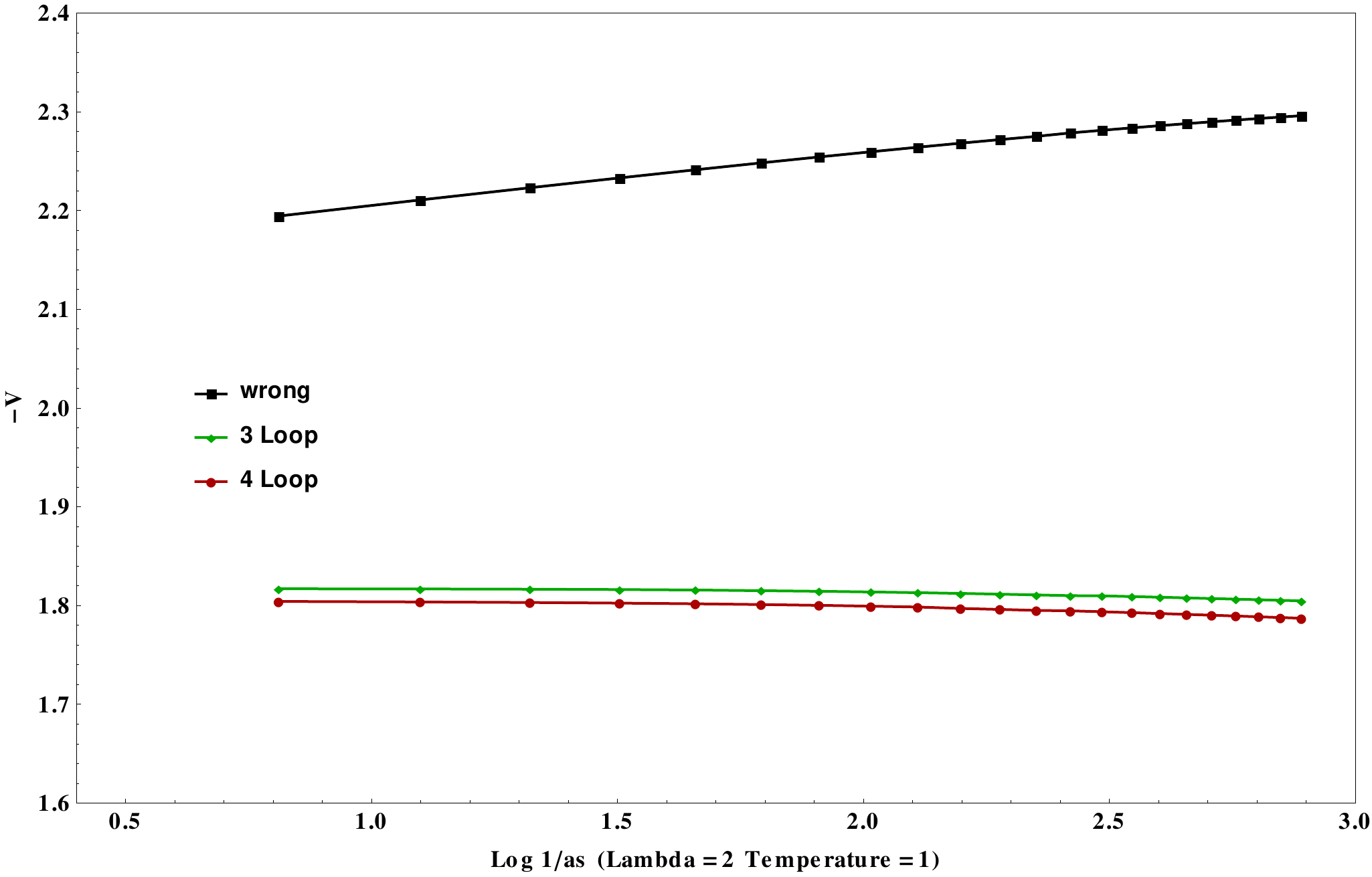}
\end{center}
\caption{The vertex -$V(0,0,0,0)$ versus ${\rm log}(1/a_s)$. The boxes (black) symbols are the 3 loop calculation with the renormalization performed incorrectly, using an extra counterterm on the basketball diagram (see text for further explanation). The diamonds (green) and circles (red) are the 3 loop and 4 loop calculations with 
the renormalization done correctly. \label{Mlogas}}
\end{figure}

\section{Discussion and Conclusions}
\label{conclusions-section}

There is a hierarchial relationship between the order of the truncation and the number of variational vertices that can be included \cite{berges-hierarchy}.
If the effective action is truncated at $L$ loops in the skeleton expansion, the corresponding $n$PI effective actions are identical for $n\ge L$. 
In this sense, a 3 loop calculation done within the 3PI formalism, a 4 loop calculation done within the 4PI formalism, etc, is complete. 
It is equivalent to say that one necessarily works with $L\ge n$. 
As noted in section \ref{introduction-section}, several calculations have been done with the 2PI effective action at the 2 and 3 loop level. 
Since the introduction of higher order variational vertices is numerically very difficult, 
we would like to know if we can extend these previous calculations by increasing $L$ without simultaneously increasing $n$.

Unfortunately, there is evidence that an $L$ loop calculation in the $n$PI formalism should, in general, be done with $L=n$.
In a gauge theory, it can be shown that the $n$ loop $n$PI effective action respects gauge invariance, to the order of the truncation \cite{Smit2003,Zaraket2004}. 
In particular, it is known that to calculate leading order transport coefficients in gauge theories with an $n$PI formalism, one must use the 3 loop 3PI effective action \cite{Carrington-transport-3pi}. 
In QED a 2 loop 2PI calculation (which is complete at 2 loop order according to the hierarchial relationship discussed above) found weak dependence on the gauge parameter \cite{Borsanyi-2pi}.
A recent 3 loop 2PI calculation in $SU(N)$ Higgs theory \cite{guy-2014} has found strong dependence on the gauge parameter. 

The issue of whether or not $n$PI calculations with $L>n$ are useful, has not been investigated previously in scalar theories. 
Three loop 2PI calculations have only been done in  symmetric $\phi^4$ theory, where the symmetry prevents 3-vertices and the 3PI theory reduces to the 2PI one. 
We have studied the convergence of the 2PI expansion at the 4 loop level. 
The Landau pole limits our ability to study large couplings, but the accessible range of parameters shows clearly that 4 loop contributions in the skeleton expansion become important at large coupling.
This kind of behaviour indicates that one should extend the calculation to the 4PI level. 

Higher order effective actions can be derived using a variety of methods \cite{berges-hierarchy,Carrington2004,Guo2011,Guo2012}, but solving the resulting variational equations is extremely difficult and little progress has been made. 
The calculation of scalar viscosity at next-to-leading order was formulated using a 4PI effective theory \cite{carrington-transport-4pi}.
A scalar 4PI theory was studied in 3 dimensions in \cite{Fu2013,Fu2014} and the 3PI action was used to study Yang-Mills theory in 3 dimensions in \cite{guy-2013}.
In spite of the inherent difficulties with these calculations, the results of this paper indicate that they are important at next-to-leading order, and motivate further efforts.

\end{document}